\documentclass[twocolumn]{revtex4}
\usepackage{amssymb,epsf}
\usepackage{latexsym}
\usepackage{xcolor}
\usepackage{epsfig}
\usepackage{float}

\begin{document}
\title{$H_0$ tension in Tsallis and R\'enyi statistics}
\author{N. Sadeghnezhad\footnote{nsadegh@maragheh.ac.ir}}
%\address{Research Institute for Astronomy and Astrophysics of Maragha (RIAAM), Maragheh, Iran}
\address{Research Institute for Astronomy and Astrophysics of Maragha
(RIAAM), University of Maragheh, P.O. Box 55136-553, Maragheh,
Iran}

\today
\begin{abstract}
Motivated by recent attempts to study the implications of Tsallis and R\'enyi statistics in gravitational, cosmological, and astrophysical systems, the possible relationships between $H_0$ tension and generalized statistics are explored. It is obtained that, in the light of $H_0$ tension, the energy-time uncertainty relations in Tsallis and R\'enyi statistics constrain the values of Tsallis and R\'enyi parameters. Hence, the way to find the footprints of non-extensivity in the universe is paved.
\end{abstract}

\maketitle

\section{Introduction}

Despite the pre-expectations, Hubble parameter measurements unveil a meaningful and nonignorable difference between the reports obtained from the Cosmic Microwave Background (CMB) and Type Ia Supernova (SN) which are $H_{\rm CMB}= 67.66 \pm 0.42\textmd{Km/s/Mpc}$ \cite{Planck:2018vyg} and $H_{\rm SN}= 74.03 \pm 1.42\textmd{Km/s/Mpc}$ \cite{SNeIa}, respectively. This controversial issue in the current value of the Hubble parameter is called the $H_0$ (Hubble) tension and has attracted a lot of attempts to itself and has raised the hopes of finding even new physics \cite{Planck:2013pxb, Hu:2023jqc}. Recently, it has been argued that the Heisenberg uncertainty principle (HUP), the energy-time uncertainty relation, and their constraints on the precision of observations may pave the way to eliminate the tension \cite{Capozziello:2020nyq}.

On the other hand, the quantum aspects of gravity, leading to generalized uncertainty principles (GUP) and relations, should be highlighted in the early universe and thus induce their footprint on CMB. Therefore, any CMB observation (and its accuracy) should be bounded by the generalized uncertainty relation instead of the energy-time uncertainty relation \cite{Moradpour:2022oxr}. Motivated by this expectation and the dominant role of quantum mechanics in describing SN and the current universe, a permanent difference between $H_{\rm CMB}$ and $H_{\rm SN}$ is shown to be expected, its value depending on the values of the parameters of quantum gravity \cite{Moradpour:2022oxr}. Indeed, the $H_0$ tension may be employed to bound the free parameters of quantum gravity \cite{Aghababaei:2021gxe, Moradpour:2022oxr}.

Ordinary statistical mechanics is based on the Shannon entropy whose special case is the Gibbs entropy. Therefore, extensive and additive systems are well described in this framework. Consequently, non-extensive systems such as those including long-range interactions such as gravity are proposed to be studied using generalized statistics like Tsallis and R\'enyi statistics providing generalized forms of Shannon entropy \cite{Renyi,tsallis}. On the other hand, the Bekesntein entropy is non-extensive, which strengthens the idea of using generalized statistics to study gravitational systems \cite{Tsallis:2012js, non23, non22, 5, non18, non19, Czinner:2017tjq}. In this regard, it has also been shown that quantum aspects of gravity inspire the R\'enyi entropy \cite{Moradpour:2019yiq}, Tsallis entropy \cite{Barrow:2020tzx}, and, generally deviation from Gaussian statistics \cite{Shababi:2020evc}. Correspondingly, the study of different cosmological, gravitational, and astrophysical systems in the frameworks of Tsallis and R\'enyi entropies has been intensified \cite{Tsallis:2012js, Moradpour:2020dfm, Tavayef:2018xwx, Moradpour:2018ivi, Moradpour:2016rcy, Sheykhi:2018dpn, Sheykhi:2022gzb, Abbasi:2020poo, TKG, Moradpour:2024azo, Moradpour:2017fmq, Sheykhi:2019bsh, Moradpour:2019wpj, Moradpour:2021soz, Sadeghnezhad:2021ekw, non23, non22, 5, non18, non19, Czinner:2017tjq, Moradpour:2020kss} revealing at least theoretical abilities in eliminating various problems.

Having the entropies of Tsallis and R\'enyi in hand, modifications to HUP and energy-time uncertainty relation (ETUR) are obtained \cite{Moradpour:2020kss}. On the other hand, due to the addressed deep connection between (quantum) gravity and non-extensivity as well as the dominant role of quantum gravity in the early universe, such modified forms of HUP and ETUR may be used to find the relevant accuracies (instead of HUP and GUP) and perform measurements at least for focusing on CMB. Indeed, the main goal is to investigate whether this tension is related to non-extensivity. To achieve this aim, focusing on Tsallis and R\'enyi entropies and the corresponding ETURs, relationships between the $H_0$ tension and the values of non-extensive parameters are derived and analyzed in the second and third sections, respectively. A summary is also provided in the last section.

\section{Tsallis}

The Tsallis entropy content ($S_q^T$) of a distribution including the $W$ number of states is defined as \cite{pla,Tsallis:2012js}

\begin{eqnarray}\label{200}
S_q^T=\frac{1}{1-q}\sum_{i=1}^{W}(P_i^q-P_i),
\end{eqnarray}

\noindent that reduces to $\frac{W^{1-q}-1}{1-q}$ when all states have the same probability \cite{pla,Tsallis:2012js}. For black holes with area $A$, it leads to \cite{Moradpour:2020dfm}

\begin{eqnarray}\label{20}
S_q^T=\frac{1}{1-q}[\exp\big(\frac{(1-q)A}{4}\big)-1],
\end{eqnarray}

\noindent generating new Tsallis holographic dark energy and shedding light on the way towards Kaniadakis holographic dark energy, the considerable models to justify the current universe based on relating the vacuum energy of quantum fields to the dark energy \cite{Moradpour:2020dfm}. The corresponding uncertainty principle and uncertainty relation are also obtained as \cite{Moradpour:2020kss}

\begin{eqnarray}\label{21}
\Delta x\Delta p\geq\frac{1}{2}\exp[-\frac{(\Delta
x)^2}{\sigma^2}],
\end{eqnarray}

\noindent and

\begin{eqnarray}\label{22}
\Delta t\Delta E\geq\frac{1}{2}\exp[-\frac{(\Delta
t)^2}{\sigma^2}].
\end{eqnarray}

\noindent respectively. Here, $\sigma\equiv\sqrt{\frac{1}{\delta\pi}}$ where
$\delta\equiv1-q$ and easily, HUP and ETUR are obtained when $q\rightarrow1$ \cite{Moradpour:2020kss}. Based on Eq.~(\ref{22}), for the minimum energy ($\Delta E_m$) extractable from a measurement during the time interval $\Delta t=\frac{1}{H}$, one finds

\begin{eqnarray}\label{23}
\Delta E_m\simeq\frac{H}{2}\exp[-\frac{1}{\sigma^2H^2}].
\end{eqnarray}

\noindent Considering two independent observations done on SN and CMB so that the same $\Delta E_m$ is extracted during $\Delta t_{SN}=\frac{1}{H_{SN}}$ and $\Delta t_{CMB}=\frac{1}{H_{CMB}}$, then we will reach

\begin{eqnarray}\label{24}
q_{SN}\simeq1\cdot2~q_{CMB}-157\cdot24,
\end{eqnarray}

\noindent as the relationship between the non-extensivity parameters of SN and CMB. Hence, measurements designed to detect $\Delta E_m$ from SN and CMB during time intervals $H^{-1}_{SN}$ and $H^{-1}_{CMB}$, respectively, pave the way to understand the trace of Tsallis statistics. Ordinary statistical mechanics is based on Shannon entropy (Gibbs entropy is a special case of the Shannon entropy) and therefore, it satisfies the HUP bounds. Thus, since quantum mechanics and ordinary statistical mechanics seem sufficient to study SN \cite{MartinMobberley}, we have $q_{SN}=1$ and thus $q_{CMB}\simeq131\cdot03$.

There is also another expression for entropy introduced by Tsallis defined as \cite{Tsallis:2012js,tsallis}

\begin{eqnarray}\label{25}
S_{q,\eta}=\sum_{i=1}^{W}P_i(\ln_q\frac{1}{P_i})^\eta,
\end{eqnarray}

\noindent that leads to $(\ln_q W)^\eta$ for equal probabilities \cite{Tsallis:2012js}. $\eta$ is also a free parameter and the corresponding black hole entropy ($S_{1,\eta}$) takes the form \cite{Tsallis:2012js}

\begin{eqnarray}\label{26}
S_{1,\eta}\propto A^\eta,
\end{eqnarray}

\noindent generating Tsallis holographic dark energy \cite{Tavayef:2018xwx}. For the corresponding uncertainty principle and uncertainty relation, we have \cite{Moradpour:2020kss}

\begin{eqnarray}\label{27}
&& \Delta x\Delta p\geq\frac{(\Delta x)^{2-2\eta}}{2},\\
&& \Delta t\Delta E\geq\frac{(\Delta t)^{2-2\eta}}{2},\nonumber
\end{eqnarray}

\noindent that recovers HUP and ETUR at the desired limit $\eta=1$ \cite{Moradpour:2020kss}, and finally yields

\begin{eqnarray}\label{28}
\frac{\ln H_{SN}}{\ln H_{CMB}}=1\cdot02\simeq\frac{2\eta_{CMB}-1}{2\eta_{SN}-1},
\end{eqnarray}

\noindent as a relationship used to estimate the values of $\eta_{CMB}$ and $\eta_{SN}$ in terms of each other. Similar to Eq.~(\ref{24}), it is obvious that there will be a permanent difference between $H_{SN}$ and $H_{CMB}$ provided they obey Eq.~(\ref{25}) with distinct values of $\eta$. Again, if quantum mechanics and ordinary statistics (based on Gibbs entropy) are enough to survey SN and the related measurements \cite{Capozziello:2020nyq, Moradpour:2022oxr, MartinMobberley}, then we face the HUP or equally $\eta_{SN}=1$ meaning that $\eta_{CMB}\simeq1\cdot01$.

\section{R\'enyi framework}

The R\'enyi entropy is defined as \cite{pla, Renyi}

\begin{eqnarray}\label{31}
\mathcal{S}=\frac{1}{1-q}\ln\sum_{i=1}^{W} P_i^{q},
\end{eqnarray}

\noindent forming the backbone of R\'enyi holographic dark energy model \cite{Moradpour:2018ivi}, and it is inspired by the quantum aspects of gravity \cite{Moradpour:2019yiq}. For the R\'enyi energy uncertainty relations, we have \cite{Moradpour:2019yiq}

\begin{eqnarray}\label{32}
\Delta t\Delta E\geq\frac{1}{2}\big(1+(1-q)\frac{\pi}{4}(\Delta
t)^2\big),
\end{eqnarray}

\noindent that finally leads to

\begin{eqnarray}\label{33}
&&\!\!\!\!\!\!\!\!(H_{SN}-H_{CMB})(\frac{4}{\pi}-\frac{1}{H_{CMB}H_{SN}})\simeq\frac{q_{SN}}{H_{SN}}-\frac{q_{CMB}}{H_{CMB}},\nonumber\\
&&\Rightarrow q_{SN}>\frac{H_{SN}}{H_{CMB}}q_{CMB},
\end{eqnarray}

\noindent where the last line is the result of $H_{SN}>H_{CMB}$. Moreover, since $\frac{1}{H_{CMB}H_{SN}}\ll1$, one may be leaded to $H_{SN}-H_{CMB}\approx\frac{q_{SN}}{H_{SN}}-\frac{q_{CMB}}{H_{CMB}}$. On one hand, the quantum aspects of the early universe are supposed to be stored in CMB and indeed, the physics supporting CMB and SN are different \cite{Hu:2023jqc}. Moreover, the R\'enyi entropy seems to be supported by the quantum aspects of gravity \cite{Moradpour:2019yiq}. Therefore, even if we completely analyze SN using HUP (or equally, $q_{SN}=0$ since SN is a phenomenon well described by quantum mechanics and ordinary statistical mechanics based on Shannon entropy \cite{MartinMobberley}), then we are still motivated to use Eq.~(\ref{32}) to focus on CMB meaning that we will have $q_{SN}=0$ and hence $q_{CMB}\approx H_{CMB}(H_{CMB}-H_{SN})<0$.

\section{Summary and concluding remarks}

Recently, motivated by the long-range nature of gravity, the usage of generalized statistics to study gravity and the related phenomena has been proposed leading to interesting results \cite{Tsallis:2012js, Moradpour:2020dfm, Tavayef:2018xwx, Moradpour:2018ivi, Moradpour:2016rcy, Sheykhi:2018dpn, Sheykhi:2022gzb, Abbasi:2020poo, TKG, Moradpour:2024azo, Moradpour:2017fmq, Sheykhi:2019bsh, Moradpour:2019wpj, Moradpour:2021soz, Sadeghnezhad:2021ekw}. Moreover, it seems that the quantum aspects of gravity, if there are any, as well as the non-extensivity of the Bekenstein entropy strengthen the motivation, especially in the case of Tsallis and R\'enyi formalisms \cite{non23, non22, 5, non18, non19, Czinner:2017tjq, Moradpour:2019yiq, Shababi:2020evc}. Therefore, the uncertainty principles and relations allowed in the Tsallis and R\'enyi statistics seem reasonable for studying the SN and especially the CMB observations. On the other hand, finding out the reason for the discrepancy between the current values of the Hubble parameter measured using the SN and CMB data is a challenge in contemporary physics that may be a signal for new physics and perspectives on understanding gravity \cite{Hu:2023jqc}.

Equipped with them, it is tried to show that the non-extensive aspects of CMB and SN may be traced using the related observations. The idea is based on extracting the same amount of energy during observations. The key outcomes are the bridges between the values of non-extensivity parameters of CMB and SN in different statistics. Therefore, if quantum mechanics and ordinary statistical mechanics (based on Shannon entropy whose special case is the Gibbs entropy) are enough to analyze SN, then the survey helps us estimate the values of the non-extensivity parameter of CMB in each statistics. Finally, it should be noted if the non-extensivity of gravity is proved, a permanent difference between $H_{CMB}$ and $H_{SN}$ is expected unless the values of the non-extensivity parameters of CMB and SN are the same.

%\section*{Acknowledgement}

%\subsection*{Data availability}
%Data sharing is not applicable to this article as no datasets were generated or analyzed during the
%current study.

%\subsection*{Declarations Conflict of interest}
%The authors declare no conflict
%of interest.

%%%%%%%%%%%%%%%%%%%%%%%%%%%%%%%%%%%%%%%%%%%%%%%%%%%%%%%


\begin{thebibliography}{99}

\bibitem{Planck:2018vyg}
N.~Aghanim \textit{et al.} [Planck],
``Planck 2018 results. VI. Cosmological parameters,''
Astron. Astrophys. \textbf{641}, A6 (2020)
[Erratum: Astron. Astrophys. \textbf{652}, C4 (2021)]
%doi:10.1051/0004-6361/201833910
%[arXiv:1807.06209 [astro-ph.CO]].

\bibitem{SNeIa}
A. G. Riess, S. Casertano, W. Yuan, L. M. Macri, D. Scolnic, 
%``Large Magellanic Cloud Cepheid Standards Provide a $1\%$ Foundation for the Determination of the Hubble Constant and Stronger Evidence for Physics beyond ΛCDM,``
Astrophys. J. 876, 85 (2019).

\bibitem{Planck:2013pxb}
P.~A.~R.~Ade \textit{et al.} [Planck],
``Planck 2013 results. XVI. Cosmological parameters,''
Astron. Astrophys. \textbf{571}, A16 (2014)
%doi:10.1051/0004-6361/201321591
%[arXiv:1303.5076 [astro-ph.CO]].

\bibitem{Hu:2023jqc}
J.~P.~Hu and F.~Y.~Wang,
``Hubble Tension: The Evidence of New Physics,''
Universe \textbf{9}, no.2, 94 (2023)
%doi:10.3390/universe9020094
%[arXiv:2302.05709 [astro-ph.CO]].

\bibitem{Capozziello:2020nyq}
S.~Capozziello, M.~Benetti and A.~D.~A.~M.~Spallicci,
``Addressing the cosmological $H_0$ tension by the Heisenberg uncertainty,''
Found. Phys. \textbf{50}, no.9, 893-899 (2020)
%doi:10.1007/s10701-020-00356-2
%[arXiv:2007.00462 [gr-qc]].

\bibitem{Moradpour:2022oxr}
H.~Moradpour, S.~Aghababaei, C.~Corda and N.~Sadeghnezhad,
``$H_0$ tension and uncertainty principles,''
Phys. Scripta \textbf{97}, no.5, 055008 (2022)
%doi:10.1088/1402-4896/ac6778

\bibitem{Aghababaei:2021gxe}
S.~Aghababaei, H.~Moradpour and E.~C.~Vagenas,
``Hubble tension bounds the GUP and EUP parameters,''
Eur. Phys. J. Plus \textbf{136}, no.10, 997 (2021)
%doi:10.1140/epjp/s13360-021-02007-5
%[arXiv:2109.14826 [gr-qc]].

\bibitem{Renyi}
A. Renyi, \textit{Probability Theory} (North-Holland, Amsterdam, 1970).

\bibitem{tsallis}
C. Tsallis, \textit{Introduction To Nonextensive Statistical Mechanics-Approaching A Complex World}, (Springer, New York, 2009)

\bibitem{pla} M. Masi,
``A step beyond Tsallis and R\'enyi entropies,''
Phys. Lett. A 338, 217 (2005).

\bibitem{Tsallis:2012js}
C.~Tsallis and L.~J.~L.~Cirto,
``Black hole thermodynamical entropy,''
Eur. Phys. J. C \textbf{73}, 2487 (2013)
%doi:10.1140/epjc/s10052-013-2487-6
%[arXiv:1202.2154 [cond-mat.stat-mech]].

\bibitem{non23} V.~G.~Czinner,
``Black hole entropy and the zeroth law of thermodynamics,''
Int. J. Mod. Phys. D \textbf{24}, no.09, 1542015 (2015)
%doi:10.1142/S0218271815420158

\bibitem{non22} 
V.~G.~Czinner and H.~Iguchi,
``R\'enyi Entropy and the Thermodynamic Stability of Black Holes,''
Phys. Lett. B \textbf{752}, 306-310 (2016)
%doi:10.1016/j.physletb.2015.11.061
%[arXiv:1511.06963 [gr-qc]].

\bibitem{5}
A.~Majhi, ``Non-extensive Statistical Mechanics and Black Hole Entropy From Quantum Geometry,''
Phys. Lett. B \textbf{775}, 32-36 (2017)
%doi:10.1016/j.physletb.2017.10.043
%[arXiv:1703.09355 [gr-qc]].

\bibitem{non18}
V.~G.~Czinner and H.~Iguchi,
``A Zeroth Law Compatible Model to Kerr Black Hole Thermodynamics,''
Universe \textbf{3}, no.1, 14 (2017)
%doi:10.3390/universe3010014

\bibitem{Czinner:2017tjq}
V.~G.~Czinner and H.~Iguchi,
``Thermodynamics, stability and Hawking-Page transition of Kerr black holes from R\'enyi statistics,''
Eur. Phys. J. C \textbf{77}, no.12, 892 (2017)
%doi:10.1140/epjc/s10052-017-5453-x
%[arXiv:1702.05341 [gr-qc]].

\bibitem{non19} 
N.~Komatsu,
``Cosmological model from the holographic equipartition law with a modified R\'enyi entropy,''
Eur. Phys. J. C \textbf{77}, no.4, 229 (2017)
%doi:10.1140/epjc/s10052-017-4800-2
%[arXiv:1611.04084 [gr-qc]].

\bibitem{Moradpour:2019yiq}
H.~Moradpour, C.~Corda, A.~H.~Ziaie and S.~Ghaffari,
``The extended uncertainty principle inspires the R\'enyi entropy,''
EPL \textbf{127}, no.6, 60006 (2019)
%doi:10.1209/0295-5075/127/60006
%[arXiv:1902.01703 [gr-qc]].

\bibitem{Barrow:2020tzx}
J.~D.~Barrow,
``The Area of a Rough Black Hole,''
Phys. Lett. B \textbf{808}, 135643 (2020)
%doi:10.1016/j.physletb.2020.135643
%[arXiv:2004.09444 [gr-qc]].

\bibitem{Shababi:2020evc}
H.~Shababi and K.~Ourabah,
``Non-Gaussian statistics from the generalized uncertainty principle,''
Eur. Phys. J. Plus \textbf{135}, no.9, 697 (2020)
%doi:10.1140/epjp/s13360-020-00726-9

\bibitem{Moradpour:2020kss}
H.~Moradpour, A.~H.~Ziaie and C.~Corda,
``Tsallis uncertainty,''
EPL \textbf{134}, no.2, 20003 (2021)
%doi:10.1209/0295-5075/134/20003
%[arXiv:2012.08316 [gr-qc]].

\bibitem{Moradpour:2020dfm}
H.~Moradpour, A.~H.~Ziaie and M.~Kord Zangeneh,
``Generalized entropies and corresponding holographic dark energy models,''
Eur. Phys. J. C \textbf{80}, no.8, 732 (2020)
%doi:10.1140/epjc/s10052-020-8307-x
%[arXiv:2005.06271 [gr-qc]].

\bibitem{Tavayef:2018xwx}
M.~Tavayef, A.~Sheykhi, K.~Bamba and H.~Moradpour,
``Tsallis Holographic Dark Energy,''
Phys. Lett. B \textbf{781}, 195-200 (2018)
%doi:10.1016/j.physletb.2018.04.001
%[arXiv:1804.02983 [gr-qc]].

\bibitem{Moradpour:2018ivi}
H.~Moradpour, S.~A.~Moosavi, I.~P.~Lobo, J.~P.~Morais Gra\c{c}a, A.~Jawad and I.~G.~Salako,
``Thermodynamic approach to holographic dark energy and the R\'enyi entropy,''
Eur. Phys. J. C \textbf{78}, no.10, 829 (2018)
%doi:10.1140/epjc/s10052-018-6309-8
%[arXiv:1803.02195 [physics.gen-ph]].

%\bibitem{SayahianJahromi:2018irq}
%A.~Sayahian Jahromi, S.~A.~Moosavi, H.~Moradpour, J.~P.~Morais Gra\c{c}a, I.~P.~Lobo, I.~G.~Salako and A.~Jawad,
%``Generalized entropy formalism and a new holographic dark energy model,''
%Phys. Lett. B \textbf{780}, 21-24 (2018)
%doi:10.1016/j.physletb.2018.02.052
%[arXiv:1802.07722 [gr-qc]].


%%%%%%%%%%%%%%%%%%%%%%%%%%%%%%%%%%%%%%%%%%%%%%%%%%%%%%%%%%
\bibitem{Moradpour:2016rcy}
H.~Moradpour,
``Implications, consequences and interpretations of generalized entropy in the cosmological setups,''
Int. J. Theor. Phys. \textbf{55}, no.9, 4176-4184 (2016)
%doi:10.1007/s10773-016-3043-6
%[arXiv:1601.05014 [gr-qc]].

\bibitem{Sheykhi:2018dpn}
A.~Sheykhi,
``Modified Friedmann Equations from Tsallis Entropy,''
Phys. Lett. B \textbf{785}, 118-126 (2018)
%doi:10.1016/j.physletb.2018.08.036
%[arXiv:1806.03996 [gr-qc]].

\bibitem{Sheykhi:2022gzb}
A.~Sheykhi and B.~Farsi,
``Growth of perturbations in Tsallis and Barrow cosmology,''
Eur. Phys. J. C \textbf{82}, no.12, 1111 (2022)
%doi:10.1140/epjc/s10052-022-11044-y
%[arXiv:2205.04138 [gr-qc]].

\bibitem{Abbasi:2020poo}
K.~Abbasi and S.~Gharaati,
``Tsallisian Gravity and Cosmology,''
Adv. High Energy Phys. \textbf{2020}, 9362575 (2020)
%doi:10.1155/2020/9362575

\bibitem{TKG}
H. Moradpour, M. Javaherian, E. Namvar, A. H. Ziaie,
``Gamow temperature in Tsallis and Kaniadakis statistics,'' Entropy 24, 797 (2022)

\bibitem{Moradpour:2024azo}
H.~Moradpour, M.~Javaherian, B.~Afshar and S.~Jalalzadeh,
``Tsallisian non-extensive stars,''
Physica A \textbf{636}, 129564 (2024)
%doi:10.1016/j.physa.2024.129564

\bibitem{Moradpour:2017fmq}
H.~Moradpour, A.~Sheykhi, C.~Corda and I.~G.~Salako,
``Implications of the generalized entropy formalisms on the Newtonian gravity and dynamics,''
Phys. Lett. B \textbf{783}, 82-85 (2018)
%doi:10.1016/j.physletb.2018.06.040
%[arXiv:1711.10336 [physics.gen-ph]].

\bibitem{Sheykhi:2019bsh}
A.~Sheykhi,
``New explanation for accelerated expansion and flat galactic rotation curves,''
Eur. Phys. J. C \textbf{80}, no.1, 25 (2020)
%doi:10.1140/epjc/s10052-019-7599-1
%[arXiv:1912.08693 [physics.gen-ph]].

\bibitem{Moradpour:2019wpj}
H.~Moradpour, A.~H.~Ziaie, S.~Ghaffari and F.~Feleppa,
``The generalized and extended uncertainty principles and their implications on the Jeans mass,''
Mon. Not. Roy. Astron. Soc. \textbf{488}, no.1, L69-L74 (2019)
%doi:10.1093/mnrasl/slz098
%[arXiv:1907.12940 [gr-qc]].

%%%%%%%%%%%%%%%%%%%%%%%%%%%%%%%%%%%%%%%%%%%%%%%%%%%%%%%%%%
\bibitem{Moradpour:2021soz}
H.~Moradpour, A.~H.~Ziaie, I.~P.~Lobo, J.~P.~Morais Gra\c{c}a, U.~K.~Sharma and A.~S.~Jahromi,
``The third law of thermodynamics, non-extensivity and energy definition in black hole physics,''
Mod. Phys. Lett. A \textbf{37}, no.12, 2250076 (2022)
%doi:10.1142/S0217732322500766
%[arXiv:2106.00378 [gr-qc]].

\bibitem{Sadeghnezhad:2021ekw}
N.~Sadeghnezhad,
``Entropic gravity and cosmology in Kaniadakis statistics,''
Int. J. Mod. Phys. D \textbf{32}, no.02, 2350002 (2023)
%doi:10.1142/S0218271823500025
%[arXiv:2111.13623 [gr-qc]].

\bibitem{MartinMobberley}
M.~Mobberley, \textit{Supernovae and How to Observe Them}, (Springer, 2007)

%%%%%%%%%%%%%%%%%%%%%%%%%%%%%%%%%%%%%%%%%%%%%%%%%%%%%%%

\end{thebibliography}
\end{document}